# Analysing the resilience of the European commodity production system with PyResPro, the Python Production Resilience package

Zampieri M.[1], Toreti A.[1], Ceglar A.[1], De Palma P.[2] and T. Chatzopoulos[3]

[1]European Commission – Joint Research Centre, Ispra IT

[2]Fincons SPA, Vimercate IT

[3]European Commission – Joint Research Centre, Seville ES

matteo.zampieri@ec.europa.eu

This paper presents a Python object-oriented software and code to compute the annual production resilience indicator. The annual production resilience indicator can be applied to different anthropic and natural systems such as agricultural production, natural vegetation and water resources. Here, we show an example of resilience analysis of the economic values of the agricultural production in Europe. The analysis is conducted for individual time-series in order to estimate the resilience of a single commodity and to groups of time-series in order to estimate the overall resilience of diversified production systems composed of different crops and/or different countries. The proposed software is powerful and easy to use with publicly available datasets such as the one used in this study.

## Software Metadata

| | |
|---|---|
| Current software version v1.0 | Uploaded with manuscript |
| Legal Code License | EUPL v1.2 - GPL |
| Code versioning system used | None |
| Software code languages, tools, and services used | Python3 |
| Compilation requirements, operating environments & dependencies | numpy, statsmodels.api, pandas, itertools, matplotlib |
| If available Link to developer documentation/manual | |
| Support email for questions | matteo.zampieri@ec.europa.eu |

Code Metadata

| | |
|---|---|
| Current code version v1.0 | Uploaded with manuscript |
| Legal Code License | EUPL v1.2 - GPL |
| Code versioning system used | None |
| Software code languages, tools, and services used | Python3 |
| Compilation requirements, operating environments & dependencies | numpy, pandas, csv |
| If available Link to developer documentation/manual | |
| Support email for questions | matteo.zampieri@ec.europa.eu |

## 1. Introduction

Climatological and hydrological disasters have grown by a staggering 3.5% per year since 1980 (Munich Re, NatCatSERVICE 2018, https://natcatservice.munichre.com/). Such disasters and extremes often distort fundamental physiological processes of plants thus affecting crop yields in terms of quantity and quality [1,2]. Following large-scale agroclimatic events, such as a heatwave or a drought, lower yields may also progressively give rise to sweeping socioeconomic repercussions upon seed prices, the value of crop production, farm income, and food security, to name a few. The implications of this chain of challenges for global food security have given birth to various recent studies that contemplate detrimental effects of climate extremes on crop yields [3–5] and agricultural and food systems [6,7]. In a future scenario where climate change is increasingly colliding with major stressors of the global food system, such as population growth, environmental degradation, and trade interdependence, the risk of food-supply instabilities attributable to more frequent and intense climate extremes is expected to increase [6,8–10]. For this reason, international organisations that monitor food systems are showing signs of increasing interest to understand the adaptive capacity of agricultural and food markets to extreme shocks (e.g., Chinese Ministry of Agriculture 2016 www.fao.org/china/news/detail-events/en/c/1191739; OECD/FAO 2018www.fao.org/3/I9166EN/I9166en.pdf).

Food security relies on the resilience of agricultural commodity markets to shocks; that is, on the capacity of supply and demand, with or without human intervention by the market participants, to continuously adjust to disruptions while still producing and providing human food and animal feed. In general, agricultural and food markets have many degrees of freedom to achieve resilience against climate variability and change through various channels that complement the dominant goal of yield improvement and production expansion. Seed traders sell new drought-resistant cultivars every year. Farmers buy those cultivars and also maneuver soil and irrigation management practices. Consumers diversify food patterns and reduce food loss and waste. Countries suffering damaging climate extremes temporarily lift trade barriers, such as import tariffs, to avoid skyrocketing domestic prices [6]. Net importers not only develop a variety of stock schemes to guarantee food security and supply stabilisation at the national level but also formalize international emergency reserves (e.g., APTERR) for humanitarian purposes [11]. Such 'market adjustments' occur irrespective of the origin of the disruption. Recent prominent examples of non-climate-driven disruptions that will likely keep impacting globally integrated food markets are the African Swine Fever and Covid-19 pandemics.

Inspired by recent efforts to quantify the resilience of ecosystems [12–16], this article illustrates how a robust and simple indicator of resilience can be calculated taking the example of food production systems. The underlying indicator informs us both about the degree of resistance of food production to damage as well as the degree of recovery. Therefore, it helps to assess and rank the capacity of domestic food markets in particular and society in general to cope with supply-chain disturbances attributable to production, trade and price disruptions of any origin. The indicator is flexible in many ways as it can be calculated (i) on variables that measure any dimension (e.g., production value as a proxy of economic performance, self-sufficiency as a proxy of food security), (ii) using freely available historical time-series, and (iii) on historical or even future-projected data. For this reason, such a synthesised resilience indicator can be highly relevant for national agricultural policy officials, who can use it as a tool to inform decision-making processes about the evolution of the most –or least– resilient food production systems.

The resilience of food production systems can be calculated consistently with the original definition of resilience, firstly introduced in ecology, as the ability of a system to absorb a shock without losing its function [17,18]. This can be achieved through the annual production resilience indicator [15,16]:

(1)  $R_p = \mu^2 / \sigma^2$,

where $\mu$ is the mean and $\sigma$ the standard deviation of the production time-series. The value of the indicator is inversely proportional to the risk of annual production losses, and it has been tested for annual agricultural time-series[15,16,19], vegetation primary production and water resources [20].

Real time-series often display non-stationarities and trends that hinder the computation of $R_p$ because $\mu$ and $\sigma$ are not well defined [16,19]. In this case, time-series need to be normalized in such a way to filter out the low-frequency variability and to isolate the annual production fluctuations whose amplitude and frequency can be related to resilience. This requires computing the smoothed production time-series using LOESS algorithm [21], for instance, which is used to normalize the production anomalies and standard deviation:

(2)  $P_i = loess(p_i)$,

(3)  $\pi_i = p_i / P_i$

(4)  $\sigma' = std(\pi_i)$

where the $p_i$ represents the production values of the time-series under evaluation, $\pi_i$ are the normalized time-series with respect to the baseline values, i.e. the smoothed time-series $P_i$, and $\sigma'$ is the standard deviation of the normalized time-series. The non-stationary crop production resilience indicator is simply given by the inverse squared standard deviation of the normalized anomalies [16,19]:

(5)  $R'_C = 1 / \sigma'^2$.

In case the production time-series is stationary, the baseline is constant and $R'_C$ is exactly equal to $R_C$.

In this paper, we present a simple and accessible code that computes production resilience from non-stationary time-series and some examples of analysis that can be conducted by stakeholders, policy makers, and any person interested in estimating the resilience of the agricultural production at the local, national and supra-national levels. This code can be easily adapted to any time-series of positively defined values.

## 2. Software Description

*PyResPro* is composed of the *ProSeries* class and some functions that are presented in the following frame:

```python
class ProSeries:

    # production time-series creation
    def __init__(self,name,p):
        self.name = name
        self.pro = p.copy()
        self.x = p[year]
        self.y = p[value]

    # length of the time-series
    def length(self):
        return len( list( self.x ) )

    # average production
    def mean(self):
        return np.mean( self.y )

    # smoothed time-series
    def smooth(self):
        span=min( 20. / float( self.length() ), 1. )
        x=np.array( self.x )
        y=np.array( self.y )
        z = sm.nonparametric.lowess( y, x, frac = span )
        yl = z[:,1]
        pl = self.pro.copy()
        pl[value] = yl
        return (pl)

    # normalized time-series
    def norm(self):
        yl = self.smooth()[value]
        yn = np.divide( self.y, yl )
        pn = self.pro.copy()
        pn[value] = yn
        return (pn)

    # production resilience
    def p_res(self):
        yn=self.norm()[value]
        r=np.power( np.divide( 1, np.std( yn ) ), 2)
        return(r)

    # sum time-series
    def __add__(self,other):
        proc = pd.merge( self.pro, other.pro, on=year)
        pro2 = proc.drop( year, axis = 1)
        pros = pro2.sum( axis = 1)
        pro = pd.DataFrame( { year: list(proc[year]), value: list(pros) } )
        names = self.name + ' + ' + other.name
        return ProSeries(names,pro)

    # copy time-series
    def copy(self):
        return ProSeries( self.name, self.pro.copy() )

    # anomaly correlation between time-series
    def acor(self,other):
```

```python
        a1 = self.norm()
        a2 = other.norm()
        proc = pd.merge( a1, a2, on = year )
        proc.drop( year, axis = 1, inplace = True )
        r = np.array( proc.corr() )[0,1]
        return r

    # plot
    def plot(self):
        fig = plt.plot( self.x, self.y, label = self.name )
        color = fig[-1].get_color()
        plt.plot( self.x, self.smooth()[value], label = '', color = color )

# computes diversified system resilience
def tot_res(name, tss):
    # tss is a list of ProSeries objects of a production system

    # extracts list of names
    labels = list(map(lambda x : x.name, tss))

    # computes mean individual productions
    imeans = list(map(lambda x : x.mean(), tss))

    # computes individual time-series resilience
    ip_res = list(map(lambda x : x.p_res(), tss))
    ip_len = list(map(lambda x : x.length(), tss))

    # computes progressively aggregated time-series
    atss = list(it.accumulate(tss, lambda x , y : x + y))

    # computes progressively aggregates time-series resilience
    ap_res = list(map(lambda x : x.p_res(), atss))
    ap_len = list(map(lambda x : x.length(), atss))

    # compute progressively accumulated time-series pairwise anomaly correlation (Pearson)
    pa_cor = list(x.acor(y) for x, y in zip(atss[:-1], tss[1:]))
    pa_cor.insert(0,0.)

    return labels,imeans,ip_res,ap_res,pa_cor,ip_len,ap_len

# resilience-diversity plot
def res_plot(totres,moreinfo=False, ylabel='mean production'):
    # unpack variables
    labels = totres[0]
    imeans = totres[1]
    ip_res = totres[2]
    ap_res = totres[3]
    pa_cor = totres[4]
    ip_len = totres[5]
    ap_len = totres[6]

    fig, ax = plt.subplots(figsize=(8,6))

    #plot mean productions and pairwise correlations
    color = (0., 1., 0., 1.)
    if moreinfo:
        llabels = list(it.starmap(lambda x, y, z : x + ' (' + str(y) + ',' + str(z) + ')', zip(labels,ip_len,ap_len)))
```

```python
        # provides a list of colors for correlations
        cmap = mpl.cm.get_cmap('jet')
        pa_col = list(map(lambda x : cmap((x+1)/2), pa_cor))
        pa_col[0] = color
        ax.bar(llabels,imeans, color=pa_col, label='mean prod. and paiwise corr.')
    else:
        llabels = labels
        ax.bar(llabels,imeans, color=color, label='mean production')
    ax.set_ylabel(ylabel)

    # plot production resilience
    ax2 = ax.twinx()
    ax2.plot(llabels,ip_res,'ko',label='individual resilience')
    ax2.plot(llabels,ap_res,'r-',label='aggregated resilience')

    plt.setp(ax.xaxis.get_majorticklabels(), rotation=90)
    # ask matplotlib for the plotted objects and their labels
    lines, labs = ax.get_legend_handles_labels()
    lines2, labs2 = ax2.get_legend_handles_labels()
    ax2.legend(lines + lines2, labs + labs2, loc=0)

    ax2.set_ylabel('annual production resilience')

    if moreinfo:
        fig.tight_layout(rect=[0, 0.03, 0.85, 0.95])
        ax3 = fig.add_axes([.875, 0.5, 0.025, 0.4])
        norm = mpl.colors.Normalize(vmin=-1, vmax=1)
        cb1 = mpl.colorbar.ColorbarBase(ax3, cmap=cmap, norm=norm, orientation='vertical')
        cb1.set_label('pairwise anomaly correlation')
    else:
        fig.tight_layout(rect=[0, 0.03, 1, 0.95])

    return fig
```

*ProSeries* class consists of the time-series object creation with a label name and production time-series, which is passed through a pandas *DataFrame*, and includes several methods:

- The *length* method returns the number of years of the time-series.

- The *mean* method returns the production average of the time-series.

- The *smooth* method returns the smoothed time-series through equation 2. The span parameter of LOESS smoothing corresponds to 20 years, which is hardcoded in this initial version of the software. This value has been calibrated in previous studies [10,22].

- The *norm* method normalizes the production time-series (equation 3).

- The *p-res* method computes the resilience indicator using equations 4 and 5.

- The __add__ methods allows aggregating two individual time-series. The result is a time-series defined by the sum of the production in the years when both individual time-series are defined.

- The *copy* method returns a shallow copy of the time-series.

- The *acor* method computes the correlation coefficient between two individual time-series.

- The *plot* method produces a plot of the original time-series and the smoothed one, both with the same colour.

The functions included in *PyResPro* are useful to analyse groups of time-series. Given a list of time-series, the function *tot_res* returns with all necessary information to characterize the total resilience of a diversified production system are the name, the (average) productions and the resilience of the individual time-series, the resilience of the time-series obtained summing progressively the individual time-series, the pairwise correlations between the time-series, the length of the individual and of the aggregated time-series.

The function *res_plot* produce a plot of all information computed by *tot_res* (if moreinfo is set to true) or just the average production and the resiliences of the individual and aggregated time-series (if moreinfo is left false).

## 3. Examples analyses

### 3.1 Data

The examples illustrated in this section are based of the freely available FAOSTAT data named 'Value of Agricultural Production'. The csv file downloaded from the FAOSTAT website (www.fao.org/faostat/en) selecting the values of gross production of all crop in European countries is used in the sample codes. According to the FAOSTAT database, European countries are Albania, Austria, Belarus, Belgium, Bosnia and Herzegovina, Bulgaria, Croatia, Czechia, Denmark, Estonia, Finland, France, Germany, Greece, Hungary, Ireland, Italy, Latvia, Lithuania, Luxembourg, Malta, Netherlands, North Macedonia, Norway, Poland, Portugal, Republic of Moldova, Romania, Russian Federation, Serbia, Serbia and Montenegro, Slovakia, Slovenia, Spain, Sweden, Switzerland, Ukraine, and United Kingdom.

Paraphrasing the FAOSTAT documentation (http://fenixservices.fao.org/faostat/static/documents/QV/QV_e.pdf), the value of gross production has been compiled by multiplying gross production in physical terms by output prices at farm gate. Thus, value of production measures production in monetary terms at the farm gate level. Since intermediate uses within the agricultural sector (seed and feed) have not been subtracted from production data, this value of production aggregate refers to the notion of "gross production".

The unit of measure of the downloaded data is in millions of USD, taking years 2004-2006 as a reference. This implies that joint macroeconomic variability (i.e., the variability of all local exchange rates w.r.t. the USD) is subsumed into the European time series of agricultural production values. However, the *smooth* function filters out the low frequency variability of this effect. Future analysis could consider the Eurozone only, converting the time series in EUR. For, the present paper, aiming aims at presenting some examples of PyResPro package application that can be easily replicated with publicly available data, no country selection and currency conversion is applied.

Two examples of application of the PyResPro package are given:

- the resilience estimation for a single time-series or simple combination of them (Section 3.2) and

- the resilience estimation for a diversified system production composed of different countries or different crops (Section 3.3).

**3.2 Individual and combined production resilience**

```python
import numpy as np

import pandas as pd
import statsmodels.api as sm
import csv
import matplotlib.pyplot as plt
import matplotlib as mpl
import itertools as it

# FAOSTAT data file
file = "FAOSTAT_data_2-25-2020_crops-europe.csv"
data = pd.read_csv(file,sep=',')

# relevant columns' names and unit of measure
year = 'Year'
value = 'Value'

# column names for the commodity and country selection
sel1_name = 'Item'
sel2_name = 'Area'

# select a random commodity and country
sel1 = 'Wheat'
sel2 = 'France'

# define time-series name
name = sel1 + ' - ' + sel2

# define time-series production data
pro = data[data[sel1_name]==sel1][data[sel2_name]==sel2][[year,value]]

# create the object
ts1 = ProSeries(name,pro)

# call the methods
print( ts1.name, ': time series length = ', ts1.length(), ', P-res = ', '{:.0f}'.format( ts1.p_res() ) )

ylabel = '2004-2006 million USD'
ts1.plot()
plt.legend()
plt.ylabel( ylabel )
```

The code reported above provides an example of a basic use of the methods of the *ProSeries* class for wheat in Italy, which is an important European commodity. The output of the print statement is: "`Wheat – France : time series length = 56 , P-res = 97`". The plot in Figure 1 allows a quick visual inspection of the time-series and of the anomalies with respect to the non-linear smoothing performed through the LOESS algorithm.

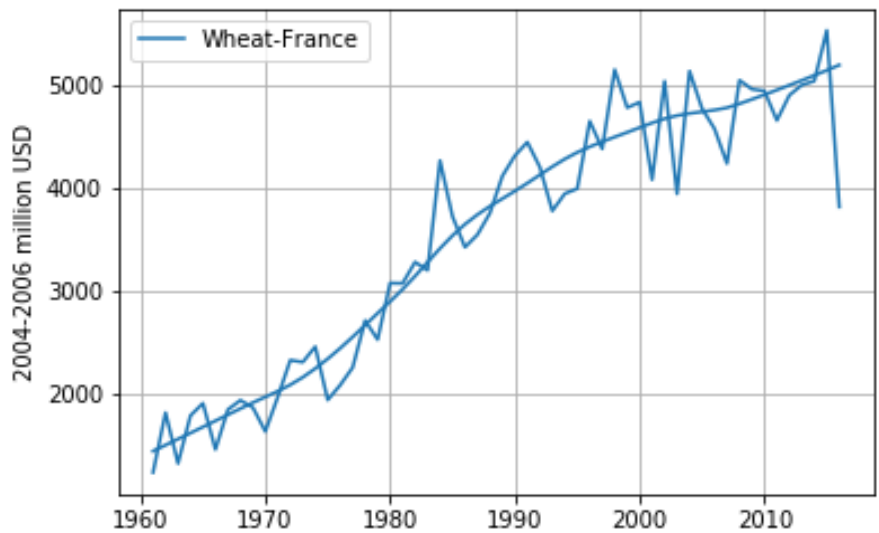

**Figure 1**: Production time series of wheat in France expressed in equivalent 2004-2006 million USD, with the smoothed time-series used for detrending. The resilience value for this time-series is $R_p$=97.

Wheat production in France increased rapidly till about the 1990s and then started stagnating [23]. This change of tendency roughly corresponds to a climatic shift bringing warmer spring temperatures in France [24,25], which shorten the growing season reducing yields [22]. Additionally, the variability increases as well in the recent decades. Worthy of notice are the effects of the 2003 heat wave [26] and the 2016 yield loss for bad weather conditions [22,27].

The same analysis can be replicated for any commodity in any country such as Italy, which is the main European durum wheat producer.

```
# select another commodity and country
sel1 = 'Wheat'
sel2 = 'Italy'

# same as before but define the time-series as "ts2"
# …
```

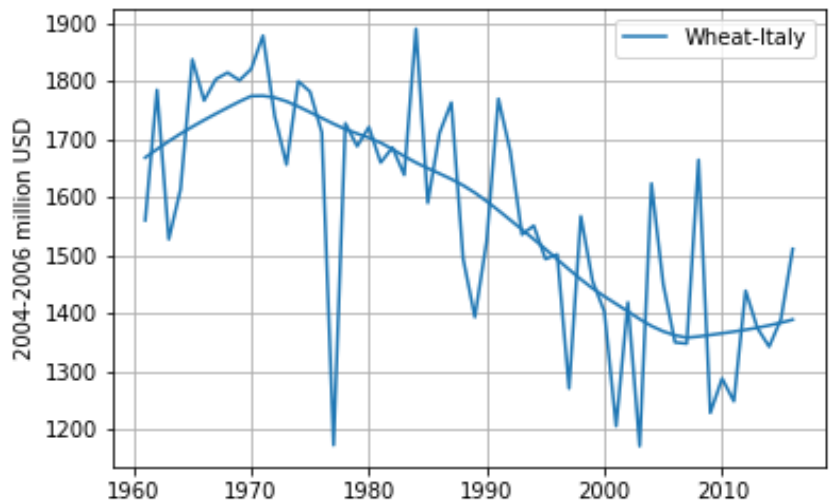

**Figure 2**: As Figure 1, but for Italy. The resilience value for this time-series is $R_p$=138.

Wheat production in Italy decreased significantly since the 1970s to 2000s. This negative trend can be attributed to reduction of sown wheat areas [19]. The largest loss happened in 1977. Large negative fluctuations are more frequent in the two most recent decades; an example of such event is year 2003, when heat wave significantly hampered growing conditions during sensitive stages in late spring. However, the fluctuations compared to the baseline values are smaller than in France, in fact the annual production resilience indicator computed for Italy is larger than that for France.

```
# sum two time-series
ts3 = ts1 + ts2
ts3.plot()
print(ts3.name,': time series length = ',ts3.length(),', P-res = ', '{:.0f}'.format(ts3.p_res()), ', correlation = ', '{:.2f}'.format(ts1.acor(ts2)))
```

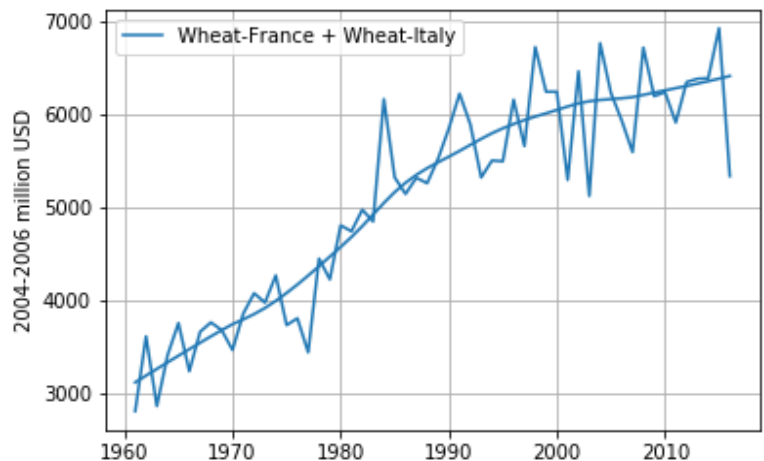

**Figure 3**: As Figure 1, but for the sum of productions in France and Italy. The resilience value for this time-series is $R_p=159$.

increasing production diversity using different crops with varying climate sensitivity to spread the risk of unfavourable climate events could be an effective measure to counteract the increase of extreme events and other negative effects of climate change. The effects of diversification on total crop production resilience can be quantified comparing the resilience of individual time series to that of the sum of production time-series in different countries or different crops. Figure 3 shows an example considering the different climates and production values of wheat in France and Italy, which is characterized by larger resilience than the individual countries. The output of the print statement gives: `Wheat-France + Wheat-Italy : time series length = 56 , P-res = 159 , correlation = 0.33`. If the two time-series were characterized by the same mean and variance, and uncorrelated between each other, the resilience of their sum is expected to double with respect to the individual ones [15,16]. The two time-series are indeed only partially correlated, with Pearson correlation coefficient equal to 0.33. The resulting compensation effect is cancelling out some of the fluctuations, despite the fact that average production in France is much larger than in Italy, also because the Italian wheat resilience is larger than the French one. The next section shows more complex analyses involving more than two time-series.

### 3.3 Diversified production system resilience

The *PyResPro* package can be used to characterize the overall production resilience of a single crop over different spatial units [19], or the resilience of a complex production system composed of different crops [16]. These are potentially more complex analyses that can be easily achieved with the methods of the *ProSeries* class and the *tot_res* and *res_plot* functions. Let's consider, for instance, wheat in the whole Europe. The code in the following frame selects the relevant data for wheat in Europe. It sorts the countries from the larger to the smaller producer, limiting the number a maximum of fifteen producers (for example) in this illustrative example. Only time-series with at least 30 production years are considered in order to achieve a sufficient accuracy in the estimation of resilience [15,16].

```python
sel1_name = 'Item'

sel2_name = 'Area'
sel1 = 'Wheat'

# select wheat data
data2 = data[data[sel1_name]==sel1]

# build up sorted list of maximum 15 time-series with at least 30 years of data
tss = []
min_length = 30
sel2_group = data2.groupby( sel2_name ).mean()
sel2_sort = sel2_group.sort_values( value, ascending=False).head(15)
for sel2 in sel2_sort.index:
    name = sel2
    pro=data[data[sel1_name]==sel1][data[sel2_name]==sel2][[year,value]]
    ts=ProSeries( name, pro )
    if ts.length() > min_length:
        tss.append( ts )
        ts.plot()
    else:
        print( sel2 )

plt.title( sel1 + ' production time-series' )
plt.ylabel( ylabel )
plt.legend( bbox_to_anchor = (1.05, 0.95) )
plt.tight_layout()
```

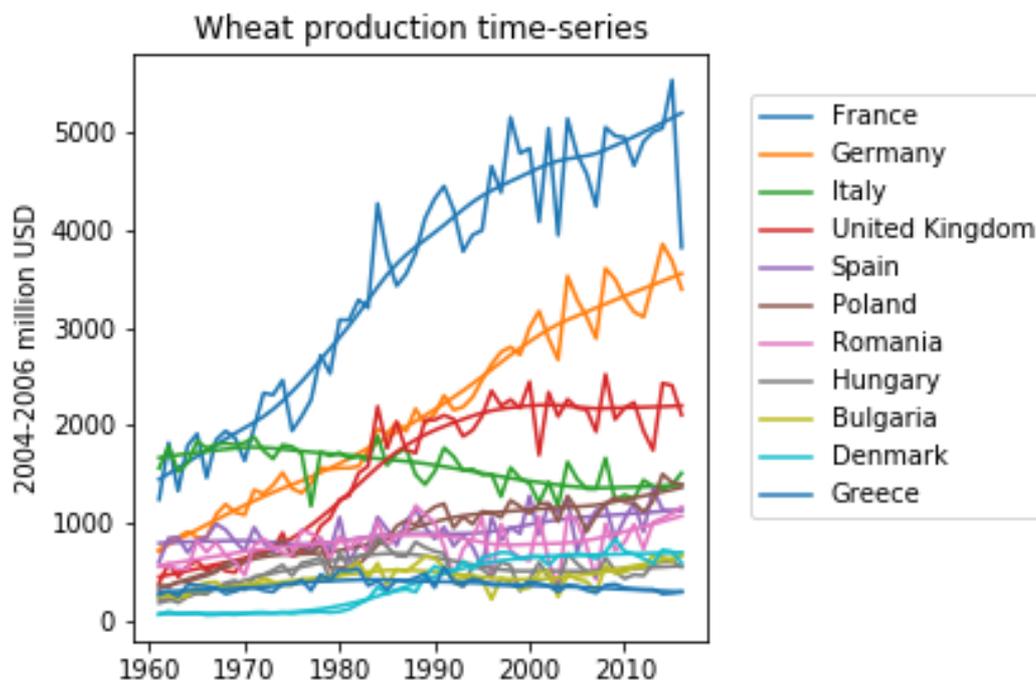

**Figure 4**: As Figure 1, but for the top European wheat producers.

The code presented above computes the resilience of the individual countries and of the progressive aggregation of countries, which is useful to understand the effect of spatial aggregation on the overall resilience of wheat production is Europe. The time-series of the top wheat producers is displayed in Figure 4. France is the most prominent producer followed by Germany, Italy and the United Kingdom.

The code in the following frame, produces the "resilience-diversity plot" (Figure 5), which is useful to understand the relative weight of the different countries in building up the overall resilience of wheat production in Europe.

```
# call the function computing system resilience
totres = tot_res(sel1, tss)

# call the function producing the resilience-diversity plot
fig = res_plot(totres, moreinfo=True, ylabel=ylabel)

fig.suptitle(sel1 + ' commodity production system resilience')
```

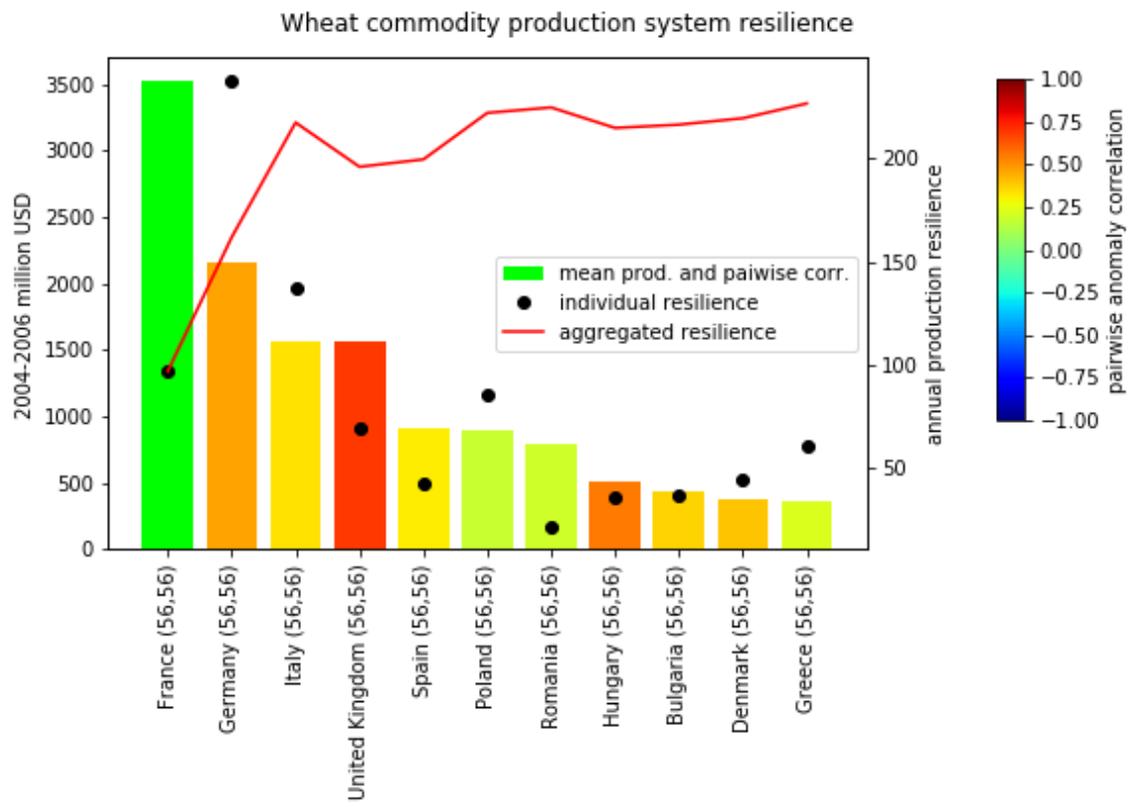

**Figure 5**: Individual (black dots) and cumulated (red lines) resilience of wheat for the top ten European countries. Cumulated resilience is obtained summing progressively the production time-series and computing the resilience from the sum of the production. For instance, cumulated resilience labelled as 'Germany' is obtained summing the production of France and Germany. The cumulated resilience labelled as 'Italy' is obtained summing the production of Italy to that of the previous two countries, and so on. The bar plot reports the average production of the different countries. The colours of the bars indicate the pairwise correlation coefficients between the production anomalies each country and the sum of the previous countries in the sorted list. Numbers in the x-labels correspond to the number of valid data in the individual time series and in the aggregated ones.

From the plot on Figure 5 it is possible to estimate the effect of the spatial aggregation of national time-series on the overall resilience of the European production system. The contribution of each country depends on the average production values. The effect of the cross-correlations between the different time series is modulating the effect of diversity on resilience [15,16]. The production resilience of the top wheat producer (France) greatly increases when Germany and Italy – i.e. the second and third producers – production are summed up. Germany is characterized by the largest individual resilience. Italian production is only partially correlated to the sum of France and Germany, which considerably increases the resilience of the aggregated time-series. The United Kingdom resilience is characterized by low resilience, and it is highly correlated to the sum of the first three countries. Thus, it does not bring positive contribution to the overall resilience. The other countries, characterized by lower individual resilience and lower production, contribute negligibly to the overall wheat production resilience.

The effects of crop diversity for the overall resilience of a single country can be conducted simply by inverting the selection labels (see code block below) and then executing that same code reported in the last three blocks presented above. This simple modification yields the results shown in Figure 6.

```
sel1_name = 'Area'
sel2_name = 'Item'
sel1 = 'Italy'

# same as previous block
# …
```

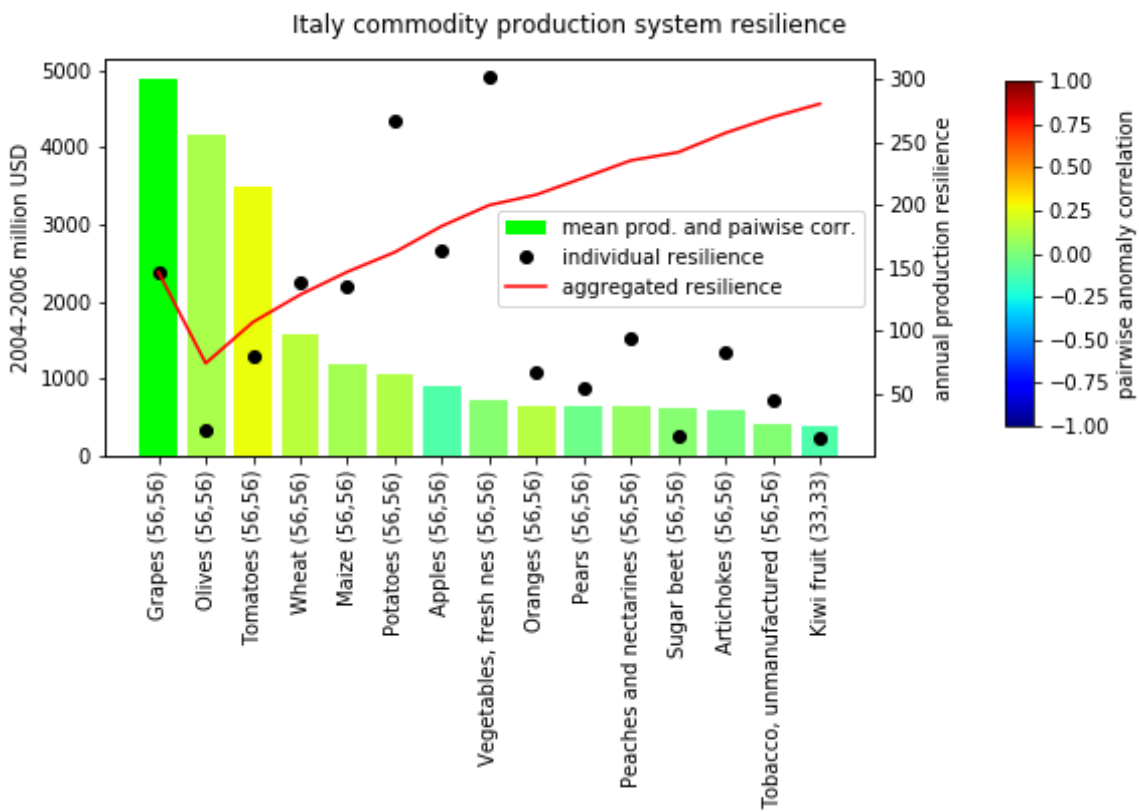

**Figure 6**: Same as Figure 5, but for the main crops in Italy.

In Italy (Figure 6), we note that olive and tomatoes are not providing additional resilience with respect to the primary commodity, which is grapes. This happens because these three crops are quite correlated between them, and olives are characterized by lower resilience. The resilience of the Italian commodity production systems gradually increases when the other commodities are considered, even though they are characterized by lower production. The accuracy of the resilience estimation decreases when Kiwi are considered, because they are represented by shorter time-series.

**5. Conclusions**

This paper presents the *PyResPro* package and provides a sensible methodology to evaluate the resilience of the commodity production system at the annual time-scale. The examples proposed in this paper are based on freely available datasets that are provided in a common structure and format, which also allow the replicability of the presented results.

The same methodology can be applied to crop production time-series evaluated in terms of raw quantity [19] or caloric content [16], but also to natural ecosystem such as vegetation primary production [20] in order to evaluate the reliability of the related ecosystem services and of water resources. The *PyResPro* software can be applied at different time-scales as well, such as the monthly or daily time-scales, and in different contexts ranging from natural science to economy and society.

We encourage the readers to use PyResPro and report any issue and ideas for future extensions, as well as requests for assistance. This first version of *PyResPro* can be copy-pasted from the present document. More developed versions can be requested directly to the corresponding author.

**References**

1. Porter, J.R.; Gawith, M. Temperatures and the growth and development of wheat: a review. *Eur. J. Agron.* **1999**, *10*, 23–36.

2. Luo, Q. Temperature thresholds and crop production: A review. *Clim. Change* 2011, *109*, 583–598.

3. Deryng, D.; Conway, D.; Ramankutty, N.; Price, J.; Warren, R. Global crop yield response to extreme heat stress under multiple climate change futures. *Environ. Res. Lett.* **2014**, *9*.

4. Powell, J.P.; Reinhard, S. Measuring the effects of extreme weather events on yields. *Weather Clim. Extrem.* **2016**, *12*, 69–79.

5. Vogel, E.; Donat, M.G.; Alexander, L. V; Meinshausen, M.; Ray, D.K.; Karoly, D.; Meinshausen, N.; Frieler, K. The effects of climate extremes on global agricultural yields. *Environ. Res. Lett.* **2019**, *14*, 54010.

6. Chatzopoulos, T.; Pérez Domínguez, I.; Zampieri, M.; Toreti, A. Climate extremes and agricultural commodity markets: A global economic analysis of regionally simulated events. *Weather Clim. Extrem.* **2019**, 100193.

7. Chen, B.; Villoria, N.B. Climate shocks, food price stability and international trade: evidence from 76 maize markets in 27 net-importing countries. *Environ. Res. Lett.* **2019**, *14*, 14007.

8. Zhu, X.; Troy, T.J. Agriculturally Relevant Climate Extremes and Their Trends in the World's Major Growing Regions. *Earth's Futur.* **2018**, *6*, 656–672.

9. Trnka, M.; Feng, S.; Semenov, M.A.; Olesen, J.E.; Kersebaum, K.C.; Rötter, R.P.; Semerádová, D.; Klem, K.; Huang, W.; Ruiz-Ramos, M.; et al. Mitigation efforts will not fully alleviate the increase in water scarcity occurrence probability in wheat-producing areas. *Sci. Adv.* **2019**, *5*, eaau2406.

10. Zampieri, M.; Ceglar, A.; Dentener, F.; Dosio, A.; Naumann, G.; van den Berg, M.; Toreti, A. When

Will Current Climate Extremes Affecting Maize Production Become the Norm? *Earth's Futur.* **2019**, *7*, 113–122.

11. Lassa, J.A.; Teng, P.; Caballero-Anthony, M.; Shrestha, M. Revisiting Emergency Food Reserve Policy and Practice under Disaster and Extreme Climate Events. *Int. J. Disaster Risk Sci.* **2019**, *10*, 1–13.

12. Kahiluoto, H.; Kaseva, J.; Balek, J.; Olesen, J.E.; Ruiz-Ramos, M.; Gobin, A.; Kersebaum, K.C.; Takáč, J.; Ruget, F.; Ferrise, R.; et al. Decline in climate resilience of European wheat. *Proc. Natl. Acad. Sci.* **2019**, *116*, 123–128.

13. Renard, D.; Tilman, D. National food production stabilized by crop diversity. *Nature* **2019**.

14. De Keersmaecker, W.; Lhermitte, S.; Honnay, O.; Farifteh, J.; Somers, B.; Coppin, P. How to measure ecosystem stability? An evaluation of the reliability of stability metrics based on remote sensing time series across the major global ecosystems. *Glob. Chang. Biol.* **2014**, *20*, 2149–2161.

15. Zampieri, M.; Weissteiner, C.; Grizzetti, B.; Toreti, A.; M., M. van den B.; Dentener, F. Estimating resilience of annual crop production systems: theory and limitations. *arXiv* **2019**.

16. Zampieri, M.; Weissteiner, C.J.; Grizzetti, B.; Toreti, A.; van den Berg, M.; Dentener, F. Estimating resilience of crop production systems: From theory to practice. *Sci. Total Environ.* **2020**, *735*, 139378.

17. Holling, C.S. Resilience and stability of ecological systems. *Annu. Rev. Ecol. Syst.* **1973**, *4*, 1–23.

18. Holling, C.S. *Engineering Resilience versus Ecological Resilience*; The National Academy of Sciences, 1996;

19. Zampieri, M.; Toreti, A.; Ceglar, A.; Naumann, G.; Turco, M.; Tebaldi, C. Climate resilience of the top ten wheat producers in the Mediterranean and the Middle East. *Reg. Environ. Chang.* **2020**, *20*, 41.

20. Zampieri, M.; Grizzetti, B.; Meroni, M.; Scoccimarro, E.; Vrieling, A.; Naumann, G.; Toreti, A. Annual Green Water Resources and Vegetation Resilience Indicators: Definitions, Mutual Relationships, and Future Climate Projections. *Remote Sens.* 2019, *11*.

21. Cleveland, W.S.; Devlin, S.J. Locally Weighted Regression: An Approach to Regression Analysis by Local Fitting. *J. Am. Stat. Assoc.* **1988**, *83*, 596–610.

22. Zampieri, M.; Ceglar, A.; Dentener, F.; Toreti, A. Wheat yield loss attributable to heat waves, drought and water excess at the global, national and subnational scales. *Environ. Res. Lett.* **2017**, *12*.

23. Schauberger, B.; Ben-Ari, T.; Makowski, D.; Kato, T.; Kato, H.; Ciais, P. Yield trends, variability and stagnation analysis of major crops in France over more than a century. *Sci. Rep.* **2018**, *8*, 16865.

24. Zampieri, M.; Scoccimarro, E.; Gualdi, S. Atlantic influence on spring snowfall over the Alps in the past 150 years. *Environ. Res. Lett.* **2013**, *8*.

25. Zampieri, M.; Toreti, A.; Schindler, A.; Scoccimarro, E.; Gualdi, S. Atlantic multi-decadal oscillation influence on weather regimes over Europe and the Mediterranean in spring and summer. *Glob. Planet. Change* **2017**, *151*, 92–100.

26. Zampieri, M.; D'Andrea, F.; Vautard, R.; Ciais, P.; De Noblet-Ducoudré, N.; Yiou, P. Hot European summers and the role of soil moisture in the propagation of mediterranean drought. *J. Clim.* **2009**, *22*.

27. Ben-Ari, T.; Boé, J.; Ciais, P.; Lecerf, R.; Van der Velde, M.; Makowski, D. Causes and implications of the unforeseen 2016 extreme yield loss in the breadbasket of France. *Nat. Commun.* **2018**, *9*, 1627.